# A Novel Online Incremental Learning Intrusion Prevention System

Christos Constantinides
School of Pure and Applied Sciences
Open University of Cyprus
Nicosia, Cyprus
christos.constantinides@st.ouc.ac.cy

Stavros Shiaeles, Bogdan Ghita
School of Computing, Electronics and Mathematics
University of Plymouth
Plymouth, UK
stavros.shiaeles@plymouth.ac.uk,
bogdan.ghita@plymouth.ac.uk

Nicholas Kolokotronis
Department of Informatics and Telecommunications
University of Peloponnese
Tripolis, Greece
nkolok@uop.gr

*Abstract*— Attack vectors are continuously evolving in order to evade Intrusion Detection systems. Internet of Things (IoT) environments, while beneficial for the IT ecosystem, suffer from inherent hardware limitations, which restrict their ability to implement comprehensive security measures and increase their exposure to vulnerability attacks. This paper proposes a novel Network Intrusion Prevention System that utilises a Self-Organizing Incremental Neural Network along with a Support Vector Machine. Due to its structure, the proposed system provides a security solution that does not rely on signatures or rules and is capable to mitigate known and unknown attacks in real-time with high accuracy. Based on our experimental results with the NSL KDD dataset, the proposed framework can achieve on-line updated incremental learning, making it suitable for efficient and scalable industrial applications.

## I. Introduction

Commercial intrusion detection/prevention systems typically suffer from low detection rates and high false positives which require substantial optimization and network specific fine tuning. The majority of these systems rely on signatures to detect potential attacks and, therefore, cannot detect "zero day", unknown attacks. Despite the significant research effort to introduce intelligence to NIDS/NIPS by implementing anomaly based/machine learning detection methods, there has been little real-world adoption of such systems. A machine learning-based NIDS/NIPS partially addresses the above inherent limitation that signature- and rule-based industrial NIDS/NIPS suffer from. Beyond the lack of real-world machine learning-based applications, there has been very limited academic work focusing on incremental learning algorithms applied to NIDS/NIPS. Incremental learning algorithms allow the classifier to refine and improve its capabilities over time (as it processes increasing amounts of input data) in contrast to an offline or batch learning algorithm, where the classifier is assumed to be exposed to the input data in a batch. Network data dynamics change significantly over time and applying static learned models progressively degrades the detection performance, making offline algorithms unsuitable for a NIDS/NIPS. We propose a novel Network Intrusion Prevention System that exploits the benefits of incremental machine learning frameworks and achieves accurate results, comparable to most offline neural network-based NIDS. The framework is based on a modified version of Self Organizing Incremental Neural Network (SOINN) [1] to achieve on-line clustering, coupled with multiple SVMs to perform classification. The evaluation of the resulting framework was performed on the NSL-KDD dataset [2], which is an improved version of well-known KDD'99 dataset. In spite of its age, the dataset is still the de-facto alternative for benchmarking techniques and tools that aim to provide effective intrusion detection systems [2]. Given its widespread use, which makes it easier to provide a reference analysis, we have also adopted this dataset to initially test our method. The results show that the proposed framework can achieve on-line updated incremental learning in a fast and efficient manner. The rest of the paper is organized as follows: the related work is discussed in Section 2, then Section 3 presents the proposed method. Section 4 gives an overview of the experimental results. Section 5 discusses the experimental results and Section 6 the provides a set of conclusions and future work to extend the proposed framework.

## II. Related works

Intrusion Detection Systems can be categorized based on the detection method used. According to Liao et al [3], detection methodologies are classified in two categories: signature-based and anomaly-based. The signature-based detection methodology is using patterns previously defined by known attacks and is distributed as a set of signatures. The signatures are then compared against patterns found in the network traffic to discover possible attacks. While effective for known threats, such a method cannot discover or prevent unknown attacks and is unable to maintain and update signatures for known or newly discovered attacks. In contrast, anomaly-based detection typically establishes a baseline/normal level using statistically significant traffic. Depending on the data analysis technique used, the training/testing process can use either classification or clustering. There has been substantial work and research towards improving the classification techniques for Intrusion Detection Systems, mostly focusing on evaluating alternative solutions for the core analysis, including neural networks [4] [5], fuzzy logic [5] [6], genetic algorithms [8], and support vector machines [9]. At the same time, clustering techniques mainly exploited the k-means and outlier detection algorithms [10], [11]. As explored in our proposed approach, prior research demonstrated the benefits of combining both stages in a hybrid approach. In their papers [2] [12] combined anomaly and misuse detection to propose a novel solution for an IDS. Following a different combination, the authors in [13] extended a neural based technique with SVMs to propose a solution for an IDS that could recognize both anomalies and known intrusions. Artificial Neural Networks were also evaluated as a possible analysis tool for Intrusion Detection Systems; in [13], the authors proposed a user behaviour model where each user is represented by a neural network, essentially a complement of a statistical model. Later, in [14], the authors used the back propagation algorithm to train the neural network for an IDS proposal with a 96% detection accuracy and a 7% false alarm; Zhang et al, 2001 in [15] built a Hierarchical Network Intrusion Detection system using statistical pre-processing and neural network classification. Following from prior studies, [16] showed that Multilayer Perceptron neural networks used for offline analysis could be applied in intrusion detection not only to recognise attacks but also to classify types of attacks. This approach, using two hidden layers, led to a 91% accuracy. Authors in [17] used a distributed time delay neural network to solve a similar multi class problem but with improved accuracy results of 97.24%. Xiang et al used a modification of SOINN [2] and applied it to intrusion detection [18]. With a semi-supervised learning-based IDS proposal, they demonstrated that the user input could be minimized by combining a modified SOINN and SVM to achieve semi-supervised learning with the same space efficiency as a supervised SVM.

A number of more recent studies extended the SOINN approach for a range of IDS-related tasks. To begin with, in [19], the authors proposed an Improved Incremental Learning SOINN for cloud-based environments. While the work aimed to detect functionality anomalies, it also provided a starting point for this work. SOINN can also be successfully applied to visualisation, as demonstrated in [20]. SOINN was also tested on an older dataset (KDD 99) in [21] and led to relatively good accuracy with only minimal pre-processing and feature extraction. On the implementation side, [22] demonstrated the efficiency of using TensorFlow towards analysing the NSL-KDD dataset and detecting malicious traffic with 96% accuracy.

As highlighted by prior research, SOINN and Incremental Learning are indeed very effective methods to tackle the challenges of intrusion detection. We aim in this paper to extend prior work by using incremental machine learning to overcome the issues from using static (offline) machine learning models applied to Network Intrusion Detection/Prevention Systems designs.

## III. PROPOSED METHOD

Throughout the evolution of IDS, it became apparent that a hybrid approach leads to better results due to its ability to discriminate using initially statistical-based approaches to improve the formulation and formatting of inputs, then apply them onto an artificial intelligence-based engine. Based on this approach, this paper proposes a conceptual framework designed to monitor inline the observed traffic and process the raw traffic to extract a set of statistical parameters that can then be passed through a detection engine. Once the detection engine determines whether the traffic is part of an attack, the packets can be logged/dropped or passed to the recipient devices. The following subsections present the overall architecture and encompassing blocks.

### A. n-SOINN-WTA-SVM

The framework, as seen in Fig. 1, consists of a detection engine that represents the core of the framework, a preprocessing module for the incoming traffic that feeds the detection engine, a validation module to evaluate the results of the detection engine and an update module that feeds back the failed results to the detection engine

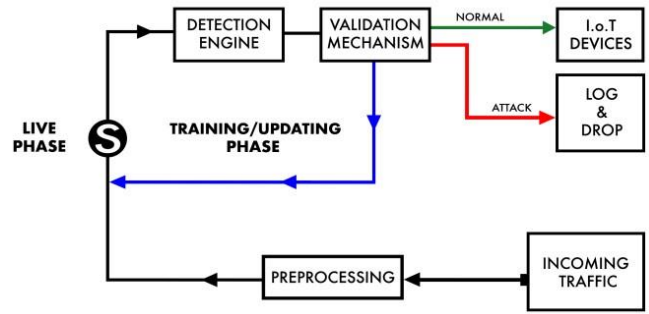

FIG. 1: MACHINE LEARNING NETWORK INTRUSION PREVENTION SYSTEM

At the core of the architecture is the detection engine, named "n-SOINN-WTA-SVM". The pre-processing module captures, extracts, and calculates the TCP-related parameters that are used as inputs for the detection engine. The validation module evaluates (either manually or automatically) the results of the detection engine and saves them in order to feed the detection engine in the update phase, whether correct or incorrect. After the initial training, the system switches between two different phases: the live phase, where the input traffic is categorized as normal or attack, allowing the system to take appropriate action, and the updating phase, where the system is incrementally updating its machine learning capabilities by learning from new input data, more specifically failed classifications. Our dataset included four different categories of attacks fed to the detection engine during experimentation and evaluation.

*1) Concept.*

The core concept of the proposed framework is to incrementally build a network protection mechanism. In the initial phase, the detection mechanism learns using a relatively small sample of network data, sufficient for basic network protection. Subsequently, as more network data becomes available, the mechanism is incrementally updated with input data of classes that it failed to detect, in order to refine and advance its protective capabilities. To support the learning process, the validation mechanism decides whether a decision failed or not. In order to advance its capabilities, the core detection engine must be able to categorize the network data in a multi-class manner, not just whether a connection is an attack or not, but what kind of attack that is and therefore provide a solution to an incremental learning multi-class problem.

*2) SOINN.*

The self-organizing and incremental neural network (SOINN) [1] is an on-line unsupervised learning mechanism for unlabelled data. SOINN has already been used in other studies as a clustering method that handles supervised data. As an online incremental clustering method, SOINN offers relatively high computational speed with low computational cost. Furthermore, the network complexity and size of SOINN are controlled and stabilised through a "garbage collector" technique. The technique defines an attrition parameter called age which represents the time after which period nodes will be removed if they are not updated for a specified time, and thus dynamically eliminate noise in the data. This property makes it attractive for dynamic environments where long-term learning is required. To ensure its scalability when expanding, the size and growth of the network are controlled by a parameter *n* where multiple pairs of SOINNs will be used as a supervised clustering method.

SOINN, as summarised in Fig. 2, initializes the network with an empty set of nodes, then adds the first two nodes to the list, with weight vectors set as the two input vectors. After the initialization, the neural network finds for every input vector the nearest node (winner) and the second-nearest node (second winner) of the input vector by measuring the distance S1 and S2 from every input to every node with the equations (1) and (2):

$$s_1 = argmin_{c \in A} dist(x, w_c) \quad (1)$$
$$s_2 = argmin_{c \in A-\{s_1\}} dist(x, w_c) \quad (2)$$

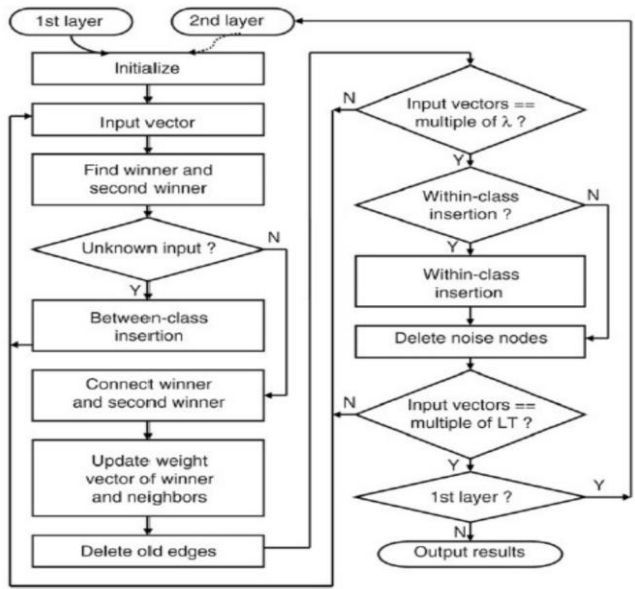

Fig. 2: SOINN Algorithm Flowchart (based on [1])

If the input vector belongs to the same cluster as the winner or second winner based on the distances calculated against a similarity threshold criterion, SOINN updates the weight vector of the node and its neighbours with the weight vector of the input vector and connects it to the node by an edge. If the input vector does not belong to the same cluster of the winner or second winner, the mechanism adds a new node to the network.

*3) n-SOINN.*
The framework proposed by this paper exploits the concept behind n-SOINN [23], which modified the original SOINN to utilize multiple pairs of SOINNs using a supervised clustering approach. n-SOINN uses two significant modifications: a global parameter that controls the topology of the network and uses the squared (rather than the basic) Euclidean distance to calculate the distance between the input and the nodes. In order to control the number of output nodes of the network, where a difference of how accurate the compressed information will be created, a parameter named *n* is introduced. This parameter dictates that any first winner node that wins more than *n* times assign a win to the second winner node. If the second winner node also has more than n winning times, a new node is generated. For n=0, the network behaves exactly like the original SOINN. Setting *n* to a very high values reduces the number of created, stable nodes and favours only the popular ones, which is likely to lead to poorer accuracy. The Euclidean distance used in the original SOINN was intended for the purpose of a single SOINN to realize the unsupervised incremental learning task. The use of the squared Euclidian distance in n-SOINN allows measurement of the distance between nodes for different SOINNs; the aging built in the SOINNs garbage collector allows eliminating nodes of the input data whenever they become unpopular.

*4) WTA-SVM.*
A Support Vector Machine (SVM) is a supervised machine learning model that operates as a group classifier by constructing a hyperplane or set of hyperplanes in a high dimensional space using training data to separate data into two groups. One of the main advantages of SVM is its effectiveness in high dimensional spaces and its ability to discriminate data which are not readily separable by simpler methods. If the 'normal' and 'abnormal' data sets to be discriminated are not linearly separable or the variations of features for two classes overlap, the SVM maps the original data space into much higher space to make the separation easier. The function for mapping of features and the other parameters should be optimized with different optimization methods.

An SVM acts as a binary classification algorithm that solves binary problems. In order to solve a multi-class problem with SVMs, the common approach is to reduce the single multiclass problem to a set of multiple binary classification problems. Two common methods are employed to solve the problem: the *one-versus-all* method, using the winner-takes-all strategy, in which the classifier with the highest output function gets the class assignment, and the *one-versus-one* method using a max-wins voting strategy, in which the class with the highest number of votes in every win of the two class problem gets the class assignment. This paper uses the former approach (one-versus-all) method and the winner-takes-all (WTA) strategy is employed to solve the multi-class problem.

*5) n-SOINN-WTA-SVM.*
The detection engine and the core of the framework, as seen in Fig. 3, is initialized with a dataset of k attack classes where *x* is a d-dimensional TCP feature vector of the connection and *y* is an attack class category label. A network connection attack category is modelled by two n-SOINNs, one with a high *n* value and another with a low *n* value. For each class *k*, a pair of two n-SOINNs and one SVM binary classifier are created. The n-SOINN with the low *n* value is supposed to hold more accurate compressed information than the n-SOINN with the high *n* value, making it a binary problem.

For every pair of n-SOINNs, the positive n-SOINN is considered to be the one with the low *n* value and the negative one with the high *n* value. The input vector of every SVM binary classifier is constructed by the output of the positive *k* n-SOINN and all other negative n-SOINNs. The output class along with its score of every binary SVM is then compared with all other binary SVMs in order to choose the top *m* classes. After choosing the top *m* classes, the output of their respective n-SOINNs is combined as an input to a multi-class SVM to get the final class.

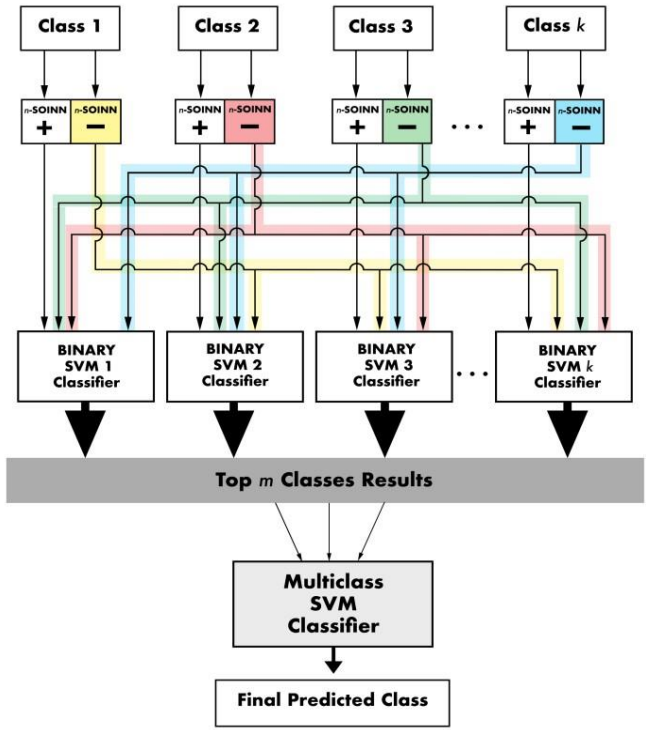

Fig. 3: n-SOINN-WTA-SVM

## B. Framework modules

### 1) Pre-processing.
This module captures and process incoming traffic in real time. The captured TCP connections are processed in order to extract the relevant features that are made available as an input vector to the detection engine - n-SOINN-WTA-SVM. For our research, the validation was performed using the 41 attributes names of NSL-KDD dataset and all attributes information are available in [2]. The framework is based on the premise that TCP characteristics of a connection suggest whether a connection is defined as an attack or not and evermore if it is defined as an attack what kind of an attack that is.

### 2) Detection Engine.
The core module of the framework consists of two main parts as seen in Fig. 3: a clustering block and a classifier. The clustering block includes a pair of n-SOINNs there are used for each class to compress the information given from the TCP connections by the pre-processing module and achieve incremental learning. The classifying part takes the output of the n-SOINNs nodes constructs an input for an SVM classifier for each class to perform the preliminary classification. Subsequently, the top *m* classes of classification pairs of each classifier sorted by score are then classified by a multiclass SVM for the final decision (*m* is user defined according to the classes available). The score is based on the distance of the samples to the separating hyperplane.

### 3) Validation Module.
The validation module serves the purpose of improving and advance the accuracy of the framework by confirming the predicted label manually or automatically. The manual method could be a user interaction in real time or automatically with the predictions saved for later use and compared against known and confirmed predictions. After confirmation, the module forwards the failed predictions to the updating module.

### 4) Updating Module.
The framework operates in two phases: ,live and updating. In the live phase, it is making decisions based on what its capabilities where at that time and an update phase where the update module is updating the system with failed predictions to improve its capabilities. The phases could run in parallel if needed in production or alternating, according to the network traffic.

## IV. EXPERIMENTAL RESULTS

Experimentation and evaluation were performed with the NSL-KDD dataset, which is an improved version of well-known KDD'99 dataset, a very good candidate dataset to evaluate the performance of any IDS and thus our framework. For the negative SOINN we used n=2 and for the positive SOINN we used n=100. As the dataset included a mixture of normal traffic and four attacks, five binary classes were created for this experiment - a normal class and four attack classes (Denial of Service, Probe, R2L and U2R). For the top *m* classes, we choose m=3 (the smallest possible number higher than 2). The implementation used LIBSVM [24], which is the best performer in a multi-class SVM out of the box; we also used the RBF kernel in both binary and multi-class SVMs.

The dataset used included two different subsets, one with 125973 records and the other one with 22544 records. The second subset is not from the same probability distribution as the first one and it also contains specific types of attacks not present in the first one. For the framework evaluation and in order to show that incremental learning is achieved we have divided the first subset into five smaller subsets used for the test/update rounds and used the second subset with 22544 records for the initial training. It should be noted that after the initial training for every update round, the subset is tested against the trained n-SOINN-WTA-SVM and only the failed predictions are fed back to the system.

TABLE I. N-SOINN-WTA-SVM RESULTS

| Round | Accuracy [%] | Time [s] | # samples |
|---|---|---|---|
| Initial training | 78.23 | 986 | 22544 |
| Round 1 | 84.44 | 1065 | 28028 |
| Round 2 | 87.98 | 1647 | 31948 |
| Round 3 | 88.88 | 2328 | 34975 |
| Round 4 | 88.96 | 2801 | 37776 |
| Round 5 | 89.67 | 3285 | 40557 |

The results, as shown in Table I, indicate that the framework can utilise incremental learning to gain substantive improvement. The samples shown in the table are the accumulated samples from the previous round. For example, for the initial training, a dataset containing 22544 records was used and for the first round of updates 5484 records was used, the number of failed predictions of the first subset out of five, accumulating the number of samples to 28028.

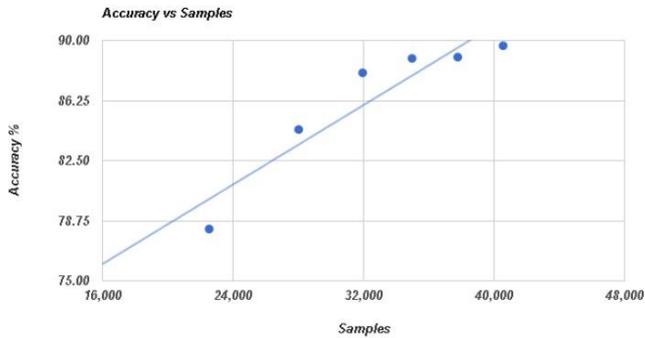
Fig. 5: Increasing Accuracy vs Sampled Feedback Chart

The accuracy chart, shown in Fig. 5, presents the relationship between the number of records accumulating and the accuracy of the system. The accuracy trend is pointing upwards with a 89.67% prediction result. The framework training samples accumulated by the end of Round 5 represented 27.30% of the total dataset; this makes it suitable for scaling applications because of its efficiency. Instead of feeding the complete dataset to the model, in our experimentation the SOINNs kept only a fraction of the information available for the whole dataset. In contrast, static models are trained with the full data available without the ability to adapt to the dynamic nature of network data over time. The time cost in seconds for the initial training was 986 seconds and after that, each update round the time cost incrementally increased with the last round costing 3285 seconds, as shown in Fig. 6.

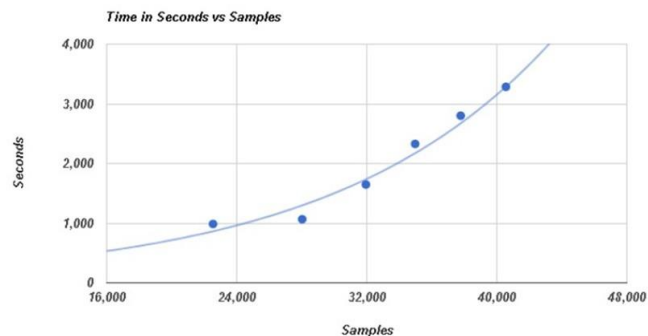
Fig. 6: Time Cost vs Samples Chart

Although the time increases over input samples of data, these are in essence rounds of updates which increase the number of feedback samples for the model. To compare the online - incremental learning method with the offline method, the framework was trained once in one batch with the same amount of input sample data. The results were 82.59% accuracy classification with a time cost of 2857 seconds.

The fact that the offline method achieved less accurate results only make the case stronger that the incremental learning roposed framework could be a competitive candidate for an industrial application.

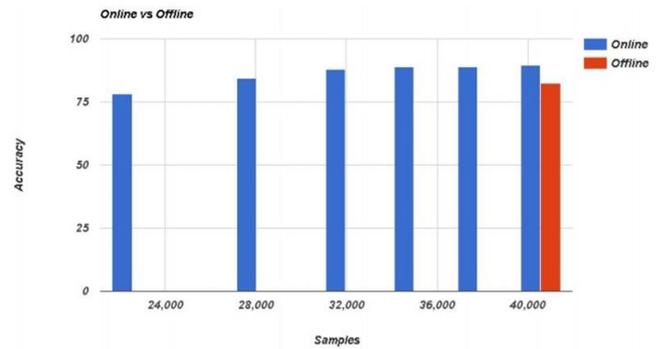
Fig. 7: Online vs Offline Method Chart

## V. DISCUSSION

The incremental learning property of the proposed framework and the evaluation results indicate that our proposal can adapt to the dynamic profile nature of network data for both normal and attacks categories. The framework utilizes less resources, is faster and it has higher detection rate than the offline method. By only feeding the system with failed predictions, not only we achieve incremental learning with promising accuracy but the framework is also efficient, meaning that, instead of feeding it the complete dataset, the SOINNs kept only a fraction of the dataset, making the framework a very good candidate for scaling systems. Although the framework training time increases as the update data input grows, the framework update and live modes could either work in parallel (simultaneously) or the update mode could switch when the framework live phase is idle (i.e. no incoming data to detect).

## VI. CONCLUSION AND FUTURE WORK

The security of the Internet of Things environments is the subject of research for almost two decades now and the rate of devices connected to the internet is growing exponentially. Each connected device is a potential target for an attack and IoT connected devices, excluding smartphones tablets and computers, have limited hardware resources which makes it even harder to apply proper security measures, increasing the exposure to vulnerabilities. The initial evaluation results shown that the proposed framework can achieve on-line updated incremental learning in a fast and efficient manner, making it suitable for an incrementally improve Network Intrusion Protection System. Our next steps are to improve the performance of the system, evaluate the proposed method with more recent datasets and attacks and finally to compare our results with state-of-the-art methods in terms of performance, False Positive and False Negative detection rate. Moreover, a possible combination of the technique with the emerging epidemic-based analysis to efficiently assess the probability of infection of a device in a network[25].

ACKNOWLEDGMENT

This project has received funding from the European Union's Horizon 2020 research and innovation programme under grant agreement no. 786698. This work reflects authors' view and Agency is not responsible for any use that may be made of the information it contains.